\documentclass[11pt]{article}
\usepackage{pgf,tikz}
\usetikzlibrary{calc,arrows}
\usepackage[all]{xy}
\usepackage{amsmath}
\usepackage{amsfonts}
\usepackage{amssymb}
\usepackage{latexsym}
\usepackage{graphicx}
\usepackage{color}
\usepackage{amscd}
\usepackage{epsfig}
\usepackage{cite}
\usepackage{circuitikz}
\usepackage[perpage,symbol]{footmisc}
\usepackage{listings}
\usepackage{array}

\newtheorem{problem}{Problem}[section]
\newtheorem{definition}[problem]{Definition}
\newtheorem{lemma}[problem]{Lemma}
\newtheorem{theorem}[problem]{Theorem}

\newtheorem{corollary}[problem]{Corollary}

\title{Proof of Mining Block-chain Systems}
\author{Chunlei Liu\footnote{School of Math., Shanghai Jiao Tong Univ., Shanghai 200240, China. 714232747@qq.com}}
\date{}
\begin{document}
\maketitle
\abstract{We propose a proof of mining system. Roughly speaking, in this system the mining stake ${\rm mstak}(A)$ with discrimination index $a\in[0,1]$ of an account $A$ is defined by the formula:
$${\rm mstak}(A)=(1-a)\cdot\frac{1}{{\rm NOM}}+a\cdot\frac{{\rm NOBM}(A)}{L},$$
where $L$ is the length of the block-chain, ${\rm NOM}$ is the number of miners in the block-chain, and ${\rm NOBM}(A)$ is the number of blocks mined by $A$}

\section{\small{BLOCK-CHAINS}}
In this section we recall the notion of block-chain systems invented by Satoshi Nakamoto \cite{Na}.
\begin{definition}A public key of a key pair in a public-key cryptography system is called an account of that system. \end{definition}
\begin{definition}A function of mass 0 on a finite set of accounts of a public-key cryptography system is called a transaction of that system.\end{definition}
\begin{definition}The signed version of transaction is the digital signature of the transaction signed by accounts on which the transaction is negative.\end{definition}
\begin{definition}A block of a public-key cryptography system is a data containing a specified account, a finite set of transactions, and the signed versions of the transactions. \end{definition}
\begin{definition}The specified account in a block of a public-key cryptography system is called the miner of the block. \end{definition}
\begin{definition}Let $B$ be a block of a public-key cryptography system, and $A$ an account of that system. The balance of $A$ in $B$ is defined by the formula
$${\rm bal}(A,B)=\sum_{{\rm tx}\in {\rm Tx}(B)}{\rm tx}(A)+{\rm rwd}(A,B),$$
where ${\rm Tx}(B)$ is the set of transactions of $B$, and $${\rm rwd}(A,B)=\left\{
                 \begin{array}{ll}
                   1, & A\hbox{ is the miner of } B,\\
                   0, & \hbox{otherewise;}
                 \end{array}
               \right.
$$
 \end{definition}
\begin{definition}Let $C$ be a sequence of blocks of a public-key cryptography system, and $A$ an account of that system. The balance of $A$ in $C$ is defined by the formula
$${\rm bal}(A,C)=\sum_{B\in C}{\rm bal}(A,B).$$
 \end{definition}
\begin{definition}A block-chain in a public-key cryptography system with a hash function is a sequence of blocks in which the hash of each block is contained in the next block and in which the balance of each account is nonnegative.
 \end{definition}
\section{\small{PROOF OF WORK}}
In this section we recall the notion of proof of work block-chain systems invented by Satoshi Nakamoto \cite{Na}.
\begin{definition}Let $B$ be a block in a public-key cryptography system with a hash function, $M$ the maximum hash value, and $D$ be a positive number. If $B$ satisfies
$${\rm hash}(B)\leq \frac{M}{D},$$
then $B$ is called a PoW block of difficulty $D$ of that system.\end{definition}

\begin{definition}Let $C=(B_0,B_1,B_2,\cdots,B_m)$ be a block-chain in a public-key cryptography system with a hash function, $L$ a positive integer, and $D=(D_0,D_1,D_2,\cdots,D_{[\frac{m}{L}]})$ a sequence of positive numbers. If
$${\rm hash}(B_i)\leq \frac{M}{D_{[\frac{i}{L}]}},$$ where $M$ is the maximum hash value, then $C$ is called a PoW block-chain with period $L$ and difficulty vector $D$ of that system. \end{definition}

\begin{definition}The computing power of a CPU with respect to a hash function is the inverse of the time it completes a single hash operation.\end{definition}

\begin{definition}The computing power of an account of a block-chain system with a hash function is the sum of computing powers of all its CPU's.\end{definition}
It is easy to prove the following.
\begin{lemma}Let $T$ be the time for a set of accounts with total computing power $P$ to find a PoW block of difficulty $D$. Then
$${\rm E}(T)\approx\frac{D}{P}.$$\end{lemma}

By the law of large numbers, we have the following theorem.
\begin{theorem}Let $L$ be a large integer, $NL\leq m<NL+L$, and $C=(B_0,B_1,\cdots,B_{m-1})$ a PoW block-chain with period $L$ and difficulty vector $D=(D_0,D_1,D_2,\cdots,D_N)$.  Let $k$ be a large integer such that $m+k-1<NL+L$. Then the time for a set of accounts with total computing power $P$ to find blocks $B_m,B_{m+1},\cdots,B_{m+k-1}$ such that $(B_0,B_1,\cdots, B_{m+k-1})$ is a PoW block-chain with period $L$ and difficulty vector $D$ is approximately $\frac{k \cdot D_N}{P}$ almost surely.\end{theorem}

\begin{definition}Let $L$ be a positive integer. A PoW block-chain system of period $L$ is a public-key cryptography system with a hash function and a communication network between the accounts in which the accounts broadcast transactions, blocks and PoW block-chains of period $L$.
 \end{definition}
\begin{definition}Let $C=(B_0,B_1,\cdots,B_m)$ be a PoW block-chain with period $L$ and difficulty vector $(D_0,D_1,D_2,\cdots,D_{[\frac{m}{L}]})$. Then we call $\sum_{i=k}^{m}D_{[\frac{i}{L}]}$ the difficulty of the segment $(B_k,B_{k+1},\cdots,B_m)$.\end{definition}
Following Satoshi Nakamoto \cite{Na}, one can show that in a PoW block-chain system of period $L$ where the majority of the computing power favours the block-chain of largest difficulty, it is almost impossible for a block-chain with a difficulty less than the largest to grow to be a block-chain of largest difficulty.
\section{\small{PROOF OF MINING}}
In this section we propose the proof of mining block-chain, and prove its security.
\begin{definition}Let $C=(B_0,B_1,B_2,\cdots,B_m)$ be a block-chain in a public-key cryptography system with a hash function, and $L$ a positive integer. We set
$$C_{L,n}=(B_{nL},B_{nL+1},\cdots,B_{nL+L-1}).$$ \end{definition}
\begin{definition}Let $C=(B_0,B_1,B_2,\cdots,B_m)$ be a block-chain in a public-key cryptography system with a hash function, and $L$ a positive integer. The number of blocks mined by an account $A$ in $C_{L,n}$ is
$${\rm NOBM}(A,C_{L,n})=\sum_{nL\leq i<nL+L}{\rm rwd}(A,B_i).$$ \end{definition}
\begin{definition}Let $C$ be a block-chain in a public-key cryptography system with a hash function, and $L$ a positive integer. The number of miners in $C_{L,n}$ is:
$${\rm NOM}(C_{L,n})=|\{A:{\rm NOBM}(A,C_{L,n})>0\}|.$$ \end{definition}
\begin{definition}Let $C=(B_0,B_1,B_2,\cdots,B_m)$ be a block-chain in a public-key cryptography system with a hash function, $L$ a positive integer, and $a\in[0,1]$. We define the mining-stake of an account $A$ in $C_{L,n}$ with discrimination index $a$ by the formula:
$${\rm mstak}(A,C_{L,n},a)=(1-a)\cdot\frac{1}{{\rm NOM}(C_{L,n})}+a\cdot\frac{{\rm NOBM}(A,C_{L,n})}{L}.$$ \end{definition}
\begin{definition}Let $C=(B_0,B_1,B_2,\cdots,B_m)$ be a block-chain in a public-key cryptography system with a hash function, $L$ a positive integer, and $D=(D_0,D_1,D_2,\cdots,D_{[\frac{m}{L}]})$ a sequence of positive numbers. If
$${\rm hash}(B_i)\leq \frac{M}{D_{[\frac{i}{L}]}}\times{\rm mstak}(A,C_{L,[\frac{i}{L}]-1},a),$$  where $M$ is the maximum hash value and $A$ is the miner of $B_i$, then $C$ is called a PoM block-chain with period $L$, difficulty vector $D$, and discrimination index $a$. \end{definition}It is easy to prove the following.
\begin{lemma}$L$ be a positive integer, $D=(D_0,D_1,D_2,\cdots,D_N)$ a sequence of positive numbers, and $a\in[0,1]$. Let $NL\leq m<NL+L$, and $C=(B_0,B_1,B_2,\cdots,B_{m-1})$ a PoM block-chain with period $L$, difficulty vector $D$, and discrimination index $a$. Let $T$ be the time for a set $S$ of accounts  to find a block $B_m$ such that $(B_0,B_1,\cdots,B_m)$ is a PoM block-chain with period $L$, difficulty vector $D$, and discrimination index $a$. Then
$${\rm E}(T)\approx\frac{ D_N}{{\rm mstak}(S,C_{L,N-1},a)},$$
where
$${\rm mstak}(S,C_{L,N-1},a)=\sum_{A\in S}{\rm mstak}(A,C_{L,N-1},a)$$\end{lemma}
By the law of large numbers, we have the following theorem.
\begin{theorem}Let $L$ be a positive integer, $D=(D_0,D_1,D_2,\cdots,D_N)$ a sequence of positive numbers, and $a\in[0,1]$. Let $NL\leq m<NL+L$, and $C=(B_0,B_1,B_2,\cdots,B_{m-1})$ a PoM block-chain with period $L$, difficulty vector $D$, and discrimination index $a$. Let $k$ be a large integer such that $m+k-1<NL+L$. Then the time for a set $S$ of accounts  to find block $B_m,B_{m+1},\cdots,B_{m+k-1}$ such that $(B_0,B_1,\cdots,B_{m+k-1})$ is a PoM block-chain with period $L$, difficulty vector $D$, and discrimination index $a$ is approximately
$$\frac{k\cdot D_N}{{\rm mstak}(S,C_{L,N-1},a)}$$
almost surely.\end{theorem}
It is easy to prove the following.
\begin{lemma}Let $C=(B_0,B_1,B_2,\cdots,B_{m-1})$ be a PoM block-chain with period $L$, and discrimination index $a\neq0$. Let $S$ be a set of accounts. Then $${\rm mstak}(S,C_{L,n},a)>\frac{1}{2}$$
if and only if
$$\frac{{\rm NOBM}(S,C_{L,n})}{L}>\frac{1-a}{a}(\frac{1}{2(1-a)}-\frac{|S|}{{\rm NOM(C_{L,n})}}),$$
where
$${\rm NOBM}(S,C_{L,n})=\sum_{A\in S}{\rm NOBM}(A,C_{L,n}a)$$ \end{lemma}
From the above lemma one can infer the following.
\begin{corollary}Let $C=(B_0,B_1,B_2,\cdots,B_{m-1})$ be a PoM block-chain with period $L$, and discrimination index $a\neq0$. Let $S$ be a set of accounts such that
$$\frac{|S|}{{\rm NOM(C_{L,n})}}\geq\frac{1}{2(1-a)}.$$ Then $${\rm mstak}(S,C_{L,n},a)>\frac{1}{2}.$$\end{corollary}

\begin{definition}Let $L$ be a positive integer, and $a\in[0,1]$. A PoM block-chain system of period $L$ and discrimination index $a$ is a public-key cryptography system with a hash function and a communication network between the accounts in which the accounts broadcast transactions, blocks and PoW block-chains of period $L$ and discrimination index $a$.\end{definition}
\begin{definition}Let $C=(B_0,B_1,\cdots,B_m)$ be a PoM block-chain with period $L$ and difficulty vector $(D_0,D_1,D_2,\cdots,D_{[\frac{m}{L}]})$. Then we call $\sum_{i=k}^{m}D_{[\frac{i}{L}]}$ the difficulty of the segment $(B_k,B_{k+1},\cdots,B_m)$.\end{definition}
Following Satoshi Nakamoto \cite{Na}, one can show that in a PoM block-chain system of period $L$ and index $a$ where the majority of the accounts favours the block-chain of largest difficulty, it is almost impossible for a block-chain with a difficulty less than the largest to grow to be a block-chain of largest difficulty.

\section{{\small CONCLUSION}}We have proposed a proof of mining block-chain system. We have shown that the proof of mining block-chain system is secure. The proof of mining block-chain system is more efficient than the proof of work system, but a litter less efficient than the
 the proof of stake block-chain systems. The proof of stake systems have been studied by many authors  \cite{KN, BGM, NXT, Mi, BPS, DGKR, KRDO, Bu, Po}.

\end{document}